\documentclass[prl,superscriptaddress,showpacs,twocolumn,preprintnumbers]{revtex4}
\usepackage{graphicx}
\usepackage{epsfig}
\usepackage{dcolumn}
\usepackage{bm}

\newcommand{\eff}{\varepsilon}
\newcommand{\BR}{{\cal B}}

\newcommand{\LK}{\mathcal{L}}

\newcommand{\psp}{\psi^\prime}
\newcommand{\psip}{\psi^\prime}
\newcommand{\jpsi}{J/\psi}
\newcommand{\EE}{e^+e^-}
\newcommand{\MM}{\mu^+\mu^-}

\newcommand{\pp}{\pi^+\pi^-}

\newcommand{\beq}{\begin{equation}}
\newcommand{\eeq}{\end{equation}}
\newcommand{\beqy}{\begin{eqnarray}}
\newcommand{\eeqy}{\end{eqnarray}}
\newcommand{\bitm}{\begin{itemize}}
\newcommand{\eitm}{\end{itemize}}

\begin{document}

\preprint{} \preprint{\vbox{ \hbox{   }
        \hbox{Belle Preprint 2012-7}
        \hbox{KEK   Preprint 2011-30}}}
\title{\quad\\[0.9cm]
Search for double charmonium decays of the $P$-wave spin-triplet
bottomonium states}

\affiliation{University of Bonn, Bonn}
\affiliation{Budker Institute of Nuclear Physics SB RAS and Novosibirsk State University, Novosibirsk 630090}
\affiliation{Faculty of Mathematics and Physics, Charles University, Prague}
\affiliation{Department of Physics, Fu Jen Catholic University, Taipei}
\affiliation{Justus-Liebig-Universit\"at Gie\ss{}en, Gie\ss{}en}
\affiliation{Gyeongsang National University, Chinju}
\affiliation{Hanyang University, Seoul}
\affiliation{University of Hawaii, Honolulu, Hawaii 96822}
\affiliation{High Energy Accelerator Research Organization (KEK), Tsukuba}
\affiliation{Indian Institute of Technology Guwahati, Guwahati}
\affiliation{Indian Institute of Technology Madras, Madras}
\affiliation{Institute of High Energy Physics, Chinese Academy of Sciences, Beijing}
\affiliation{Institute of High Energy Physics, Vienna}
\affiliation{Institute of High Energy Physics, Protvino}
\affiliation{INFN - Sezione di Torino, Torino}
\affiliation{Institute for Theoretical and Experimental Physics, Moscow}
\affiliation{J. Stefan Institute, Ljubljana}
\affiliation{Kanagawa University, Yokohama}
\affiliation{Institut f\"ur Experimentelle Kernphysik, Karlsruher Institut f\"ur Technologie, Karlsruhe}
\affiliation{Korea Institute of Science and Technology Information, Daejeon}
\affiliation{Korea University, Seoul}
\affiliation{Kyungpook National University, Taegu}
\affiliation{\'Ecole Polytechnique F\'ed\'erale de Lausanne (EPFL), Lausanne}
\affiliation{Faculty of Mathematics and Physics, University of Ljubljana, Ljubljana}
\affiliation{Luther College, Decorah, Iowa 52101}
\affiliation{University of Maribor, Maribor}
\affiliation{Max-Planck-Institut f\"ur Physik, M\"unchen}
\affiliation{University of Melbourne, School of Physics, Victoria 3010}
\affiliation{Graduate School of Science, Nagoya University, Nagoya}
\affiliation{Kobayashi-Maskawa Institute, Nagoya University, Nagoya}
\affiliation{Nara Women's University, Nara}
\affiliation{National Central University, Chung-li}
\affiliation{National United University, Miao Li}
\affiliation{Department of Physics, National Taiwan University, Taipei}
\affiliation{H. Niewodniczanski Institute of Nuclear Physics, Krakow}
\affiliation{Nippon Dental University, Niigata}
\affiliation{Niigata University, Niigata}
\affiliation{University of Nova Gorica, Nova Gorica}
\affiliation{Osaka City University, Osaka}
\affiliation{Pacific Northwest National Laboratory, Richland, Washington 99352}
\affiliation{Panjab University, Chandigarh}
\affiliation{Research Center for Nuclear Physics, Osaka University, Osaka}
\affiliation{University of Science and Technology of China, Hefei}
\affiliation{Seoul National University, Seoul}
\affiliation{Sungkyunkwan University, Suwon}
\affiliation{School of Physics, University of Sydney, NSW 2006}
\affiliation{Tata Institute of Fundamental Research, Mumbai}
\affiliation{Excellence Cluster Universe, Technische Universit\"at M\"unchen, Garching}
\affiliation{Toho University, Funabashi}
\affiliation{Tohoku Gakuin University, Tagajo}
\affiliation{Tohoku University, Sendai}
\affiliation{Department of Physics, University of Tokyo, Tokyo}
\affiliation{Tokyo Institute of Technology, Tokyo}
\affiliation{Tokyo Metropolitan University, Tokyo}
\affiliation{Tokyo University of Agriculture and Technology, Tokyo}
\affiliation{CNP, Virginia Polytechnic Institute and State University, Blacksburg, Virginia 24061}
\affiliation{Yamagata University, Yamagata}
\affiliation{Yonsei University, Seoul}
  \author{C.~P.~Shen}\affiliation{Graduate School of Science, Nagoya University, Nagoya} 
  \author{C.~Z.~Yuan}\affiliation{Institute of High Energy Physics, Chinese Academy of Sciences, Beijing} 
  \author{T.~Iijima}\affiliation{Kobayashi-Maskawa Institute, Nagoya University, Nagoya}\affiliation{Graduate School of Science, Nagoya University, Nagoya} 
  \author{I.~Adachi}\affiliation{High Energy Accelerator Research Organization (KEK), Tsukuba} 
  \author{H.~Aihara}\affiliation{Department of Physics, University of Tokyo, Tokyo} 
  \author{K.~Arinstein}\affiliation{Budker Institute of Nuclear Physics SB RAS and Novosibirsk State University, Novosibirsk 630090} 
  \author{D.~M.~Asner}\affiliation{Pacific Northwest National Laboratory, Richland, Washington 99352} 
  \author{T.~Aushev}\affiliation{Institute for Theoretical and Experimental Physics, Moscow} 
  \author{A.~M.~Bakich}\affiliation{School of Physics, University of Sydney, NSW 2006} 
  \author{B.~Bhuyan}\affiliation{Indian Institute of Technology Guwahati, Guwahati} 
% ...
\collaboration{The Belle Collaboration}


\begin{abstract}

Using a sample of 158 million $\Upsilon(2S)$ events collected with
the Belle detector, we search for the first time for double
charmonium decays of the $P$-wave spin-triplet bottomonium states
($\Upsilon(2S) \to \gamma\chi_{bJ}$, $\chi_{bJ} \to \jpsi \jpsi$, $\jpsi \psp$, $\psp \psp$ for $J=0$,
$1$, and $2$). No significant $\chi_{bJ}$ signal is observed in
the double charmonium mass spectra, and we obtain the following upper limits,
$\BR(\chi_{bJ} \to \jpsi
\jpsi)<7.1\times 10^{-5}$, $2.7\times 10^{-5}$, $4.5\times
10^{-5}$, $\BR(\chi_{bJ} \to \jpsi \psp)<1.2\times 10^{-4}$,
$1.7\times 10^{-5}$, $4.9\times 10^{-5}$, $\BR(\chi_{bJ} \to \psp
\psp)<3.1\times 10^{-5}$, $6.2\times 10^{-5}$, $1.6\times 10^{-5}$
for $J=0$, $1$, and $2$, respectively, at the 90\% confidence level. These
limits are significantly lower than the central values (with uncertainties of
50\% to 70\%) predicted
using the light cone formalism but are consistent with
calculations using the NRQCD factorization approach.

\end{abstract}


\maketitle


The order of magnitude discrepancy between the cross sections of
the double charmonium production processes $\EE \to \jpsi \eta_c$,
$\jpsi \eta_c'$, $\psp \eta_c$, $\psp \eta_c'$, $\jpsi \chi_{c0}$,
and $\psp \chi_{c0}$ measured in the Belle~\cite{bellec1, bellec2}
and BaBar~\cite{babarc} experiments and those of the leading-order
non-relativistic QCD (NRQCD) predictions~\cite{nrqcd1, nrqcd2,
nrqcd3} has been a great challenge to theorists. After great efforts
it was shown that agreement can be achieved by taking
into account QCD radiative and relativistic
corrections~\cite{nrqcd1,Bodwin,zhang,Bodwin2,He}.

As in $e^+e^-$ annihilation, double
charmonium final states can also be formed in bottomonium
decays, which provide a new test of the dynamics of hard
exclusive processes and charmonium structure. While the rate of
$\eta_b\to \jpsi\jpsi$ decay was calculated a long time ago by
many authors~\cite{etab}, the rates for $P$-wave spin-triplet bottomonium states $\chi_{bJ}$
($J$=0, 1, 2) decays into double charmonium states were calculated only recently
using various theoretical models after a first
attempt about 30 years ago with a perturbative QCD method~\cite{Kart}.

The authors of Ref.~\cite{zhangjuan} calculated $\chi_{bJ} \to \jpsi \jpsi$ in
the framework of the NRQCD factorization formalism, including
second-order relativistic corrections in the relative
charm-quark velocity $v_c$, as well as an electromagnetic correction.
The branching fraction is predicted to be of order $10^{-5}$ for
$\chi_{b0}$ or $\chi_{b2}\to J/\psi J/\psi$, and $10^{-11}$ for
$\chi_{b1}\to J/\psi J/\psi$. The authors of Ref.~\cite{sang}
considered corrections to all orders in $v_c$ in the
charmonium rest frame, and found decay partial widths that are
about a factor of three larger than those in
Ref.~\cite{zhangjuan}. In the light cone (LC) formalism, however,
much larger production rates (with uncertainties of 50\% to 70\%) are
obtained in Ref.~\cite{vv}: $\BR(\chi_{bJ} \to \jpsi \jpsi) = 9.6
\times 10^{-5}$ or $1.1 \times 10^{-3}$, $\BR(\chi_{bJ} \to \jpsi
\psp) = 1.6 \times 10^{-4}$ or $1.6 \times 10^{-3}$, and
$\BR(\chi_{bJ} \to \psp \psp) = 6.6 \times 10^{-5}$ or $5.9 \times
10^{-4}$ for $J$=0 or 2, respectively.
These results are not very different from the perturbative QCD
calculation with the relative motion of the charm-quark in the
$\chi_b$ decays taken into account~\cite{vv2}.
It is therefore necessary to pin
down the source of such a significant difference between various models.


In this paper, we report a search for $\chi_{bJ}$ decays to double
charmonium in $\Upsilon(2S)$ radiative transitions, $i.e.$,
$\Upsilon(2S) \to \gamma \chi_{bJ} \to \gamma \jpsi \jpsi$,
$\gamma \jpsi \psp$, and $\gamma \psp \psp$. In order to detect
the signal efficiently, we only require one $\jpsi$ or $\psp$
candidate to be fully reconstructed (or ``tagged"). We require
that the missing mass of the $\jpsi$ (or $\psp$) and the radiative
photon candidate be in the $\jpsi$ or $\psp$ mass region. The
missing mass is defined as $M_{\rm miss} =\sqrt {(P_{e^+
e^-}-P_f)^2}$, where $P_{e^+ e^-}$ is the 4-momentum of the $e^+
e^-$ collision system, and $P_f$ is the sum of the 4-momentum of
the observed final-state particles. Double counting of an event is
allowed if both charmonium states satisfy the tag criteria. The
probability of double counting depends on the final states and
varies from a few per mille to less than 5\%; this is taken into
account in the efficiency estimation using MC samples generated
with both charmonium states decaying generically. For $\gamma
\jpsi \jpsi$, we reconstruct one $\jpsi$ signal from $\ell^+
\ell^-$ ($\ell=e$ or $\mu$) and require the missing mass of
$\gamma \jpsi$ be within the $\jpsi$ mass region. For $\gamma
\jpsi \psp$, three modes are included: (1) $\jpsi \to \ell^+
\ell^-$ with $M_{\rm miss}(\gamma \jpsi$) required to be within
the $\psip$ mass region; (2) $\psip \to \pp \jpsi \to \pp \ell^+
\ell^- $ with $M_{\rm miss}(\gamma \psip$) within the $\jpsi$ mass
region;  (3) $\psip$ $\to \ell^+ \ell^- $ with $M_{\rm
miss}(\gamma \psip$) within the $\jpsi$ mass region. For $\gamma
\psp \psp$, two modes are used: (1) one $\psip \to \pp \jpsi \to
\pp \ell^+ \ell^- $ is identified and $M_{\rm miss}(\gamma \psip$)
is required to be within the $\psip$ mass region; (2)  one $\psip
\to \ell^+ \ell^-$ is identified and $M_{\rm miss}(\gamma \psip$)
is required to be within the $\psip$ mass region.

This analysis is based on a
24.7~fb$^{-1}$ $\Upsilon(2S)$ data sample (158 million
$\Upsilon(2S)$ events~\cite{y2srad}), and a 89.4~fb$^{-1}$
continuum data sample collected at $\sqrt{s}=10.52$~GeV. Here
$\sqrt{s}$ is the center-of-mass (C.M.) energy of the colliding
$e^+e^-$. The data are collected with the Belle
detector~\cite{Belle} operating at the KEKB asymmetric-energy
$\EE$ collider~\cite{KEKB}.

{\sc evtgen}~\cite{evtgen} was used to generate Monte Carlo (MC) simulation
events. For signal MC samples, the angular distribution for
$\Upsilon(2S) \to \gamma \chi_{bJ}$ is simulated assuming a pure
$\rm{E1}$ transition ($dN/d\cos\theta_{\gamma} \propto 1+ \alpha
\cos^2\theta_{\gamma}$, $\alpha=1$, $-\frac{1}{3}$, $\frac{1}{13}$
for $J=0$, 1, 2, respectively~\cite{mcang}). Here
$\theta_\gamma$ is the polar angle of the $\Upsilon(2S)$ radiative
photon in the $\EE$ C.M. frame.
Uniform phase
space is used for $\chi_{bJ}$ decays~\cite{chibjdecays}.

Since no experimental measurements are available for the widths
of $\chi_{bJ}$~\cite{PDG},
and the theoretical expectations are at 1~MeV level or less~\cite{zhangjuan},
the widths of $\chi_{bJ}$ are set to be zero. Generic decay modes are used
for the $\jpsi$ and $\psp$. $\Upsilon(2S)$ MC events with generic decays
produced with {\sc pythia}~\cite{pythia} with two times the effective
luminosity of data are used to check the possible backgrounds
from $\Upsilon(2S)$ decays.

The detector is described in detail elsewhere~\cite{Belle}. It is
a large-solid-angle magnetic spectrometer that consists of a
silicon vertex detector (SVD), a 50-layer central drift chamber
(CDC), an array of aerogel threshold Cherenkov counters (ACC),
a barrel-like arrangement of time-of-flight scintillation counters
(TOF), and an electromagnetic calorimeter comprised of CsI(Tl)
crystals (ECL) located inside a superconducting solenoid coil
that provides a 1.5~T magnetic field. An iron flux-return located
outside the coil is instrumented to detect $K_L^0$ mesons and
to identify muons (KLM).


For well reconstructed charged tracks, the impact parameters
perpendicular to and along the beam direction with respect to the
nominal interaction point are required to be less than 0.5~cm and 4~cm,
respectively, and the transverse momentum in the laboratory frame
is required to be larger than 0.1~$\hbox{GeV}/c$. We require the
number of well reconstructed charged tracks to be greater than three for
$\gamma \jpsi \jpsi$, and greater than four for $\gamma \jpsi \psp$ and $\gamma
\psp \psp$. For the modes with
$\psp$ in the final states, events with exactly four charged
tracks are removed to suppress the significant background from QED
processes.
For each charged track, information from different
detector subsystems is combined to form a likelihood
$\mathcal{L}_i$ for each particle species~\cite{pid}. A track with
$\mathcal{R}_K = \frac{\mathcal{L}_K} {\mathcal{L}_K +
\mathcal{L}_\pi}< 0.4$ is identified as a pion with an efficiency of about
97\% for the momentum range of interest; about 3.5\% are misidentified $K$ tracks.
For electron identification, the likelihood ratio is defined as $\mathcal{R}_e
= \frac{\mathcal{L}_e}{\mathcal{L}_e+\mathcal{L}_x}$, where
$\mathcal{L}_e$ and $\mathcal{L}_x$ are the likelihoods for
electron and non-electron, respectively, determined using the
ratio of the energy deposit in the ECL to the momentum measured in
the SVD and CDC, the shower shape in the ECL, matching between
the position of the charged track trajectory and the cluster position
in the ECL, the hit information from the ACC and the $dE/dx$
information in the CDC~\cite{EID}. For muon identification, the
likelihood ratio is defined as $\mathcal{R}_\mu =
\frac{\mathcal{L}_\mu} {\mathcal{L}_\mu + \mathcal{L}_\pi +
\mathcal{L}_K}$, where $\mathcal{L}_\mu$, $\mathcal{L}_\pi$, and
$\mathcal{L}_K$ are the likelihoods for muon, pion, and kaon
hypotheses, respectively, based on the matching quality and
penetration depth of associated hits in the KLM~\cite{MUID}.

A neutral cluster is used as a photon candidate
if it does not match the extrapolation of any charged track and
its energy is greater than 50~MeV.
In calculating the recoil mass of $\gamma\jpsi$ or $\gamma\psp$,
all photon candidates except those within 0.05 radians of the
electron/positron tracks are included. No $\pi^0$ signal is observed in combining the low energy
radiative photon with any of the remaining photon candidates in
the event after all the selection criteria are applied.

\begin{figure*}[htbp]
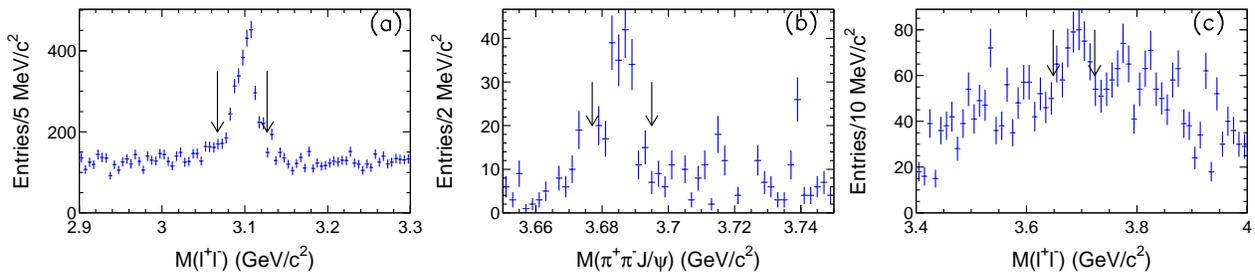

\includegraphics[height=3.5cm]{fig1a.epsi}
\includegraphics[height=3.5cm]{fig1b.epsi}
\includegraphics[height=3.5cm]{fig1c.epsi}
\caption{\label{mll-mppll} The reconstructed $\jpsi \to \ell^+
\ell^-$ (a), $\psp \to \pp \jpsi$ (b), and $\psp \to \ell^+ \ell^-$
(c) candidates mass distributions from data. The arrows show the
required signal mass regions. }
\end{figure*}

In order to correct for the effect of bremsstrahlung and
final-state radiation, photons detected in the ECL within
0.05~radians of the original $e^+$ or $e^-$ direction are included
in the calculation of the $e^+/e^-$ momentum. For the lepton pair
used to reconstruct $\jpsi$, both tracks should have
$\mathcal{R}_e>0.95$ in the $\EE$ mode; or one track should have
$\mathcal{R}_\mu>0.95$ while the other should satisfy
$\mathcal{R}_\mu>0.05$ in the $\MM$ mode.
The lepton pair identification efficiency is about 90\% for
$\jpsi\to e^+ e^-$ and 87\% for $\jpsi\to \mu^+ \mu^-$.
In order to improve the
$\jpsi$ momentum resolution, a mass-constrained fit is then
performed for $\jpsi$ signals in all the modes. As different modes
have almost the same $\jpsi$ mass resolutions, the $\jpsi$ signal
region is defined as $|M_{\ell^+\ell^-}-m_{\jpsi}| <
0.03~\hbox{GeV}/c^2$ ($\approx 2.5\sigma$), where $m_{\jpsi}$ is
the nominal mass of $\jpsi$~\cite{PDG}. The $\jpsi$ mass sidebands
are defined as $2.97~\hbox{GeV}/c^2< M_{\ell^+\ell^-} <
3.03$~GeV/$c^2$ or $3.17~\hbox{GeV}/c^2 < M_{\ell^+\ell^-} <
3.23$~GeV/$c^2$, and are twice as wide as the signal region.
 For $\psp \to \ell^+ \ell^-$, the $\psp$ signal region
is defined as $|M_{\ell^+\ell^-}-m_{\psip}|<0.0375$~GeV/$c^2$
($\approx  2.5 \sigma$), where $m_{\psip}$ is the nominal mass of
$\psp$~\cite{PDG}. The $\psip$ mass sidebands are defined as
$3.535~\hbox{GeV}/c^2 <M_{\ell^+\ell^-}<3.610$~GeV/$c^2$ or
$3.760~\hbox{GeV}/c^2 <M_{\ell^+\ell^-}<3.835$~GeV/$c^2$, and are
twice as wide as the signal region. For $\psp \to \pp \jpsi$, we
require the two pion candidates be positively identified. The
$\psp$ signal region is defined as $|M_{\pp \jpsi} -m_{\psp}| <
0.009$~GeV/$c^2$ ($\approx 3 \sigma$). Figure~\ref{mll-mppll}
shows the mass distributions of the reconstructed $\jpsi \to \ell^+
\ell^-$ (a), $\psp \to \pp \jpsi$ (b) and $\psp \to \ell^+ \ell^-$
(c) candidates.

\begin{figure*}[htbp]
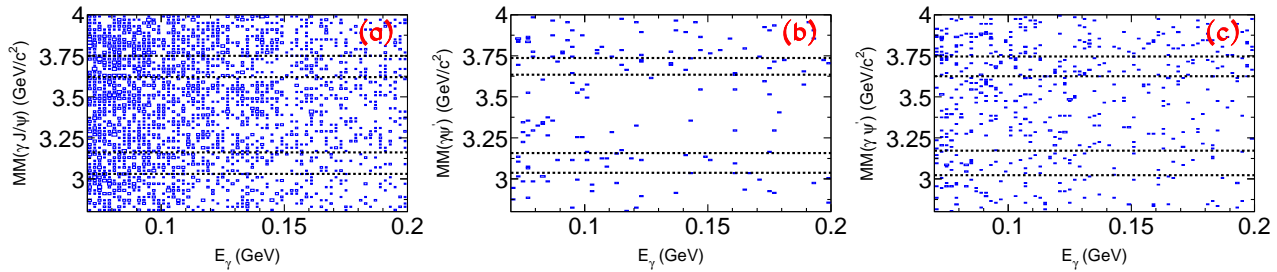

\includegraphics[height=3.5cm]{fig2a.epsi}
\includegraphics[height=3.5cm]{fig2b.epsi}
\includegraphics[height=3.5cm]{fig2c.epsi}
\caption{Scatter plots of the photon spectra in the $\EE$ C.M. frame
versus (a) $M_{\rm miss}(\gamma \jpsi)$ with $\jpsi \to \ell^+ \ell^-$
reconstructed, (b) $M_{\rm miss}(\gamma \psp)$ with $\psp \to \pp \jpsi$
reconstructed, and (c) $M_{\rm miss}(\gamma \psp)$ with $\psp \to \ell^+
\ell^-$ reconstructed. The dotted lines show the $\jpsi$ or $\psp$ signal regions ($\approx \pm3\sigma$).}
\label{llmass}
\end{figure*}

Figure~\ref{llmass} shows scatter plots
of the photon spectra in the $\EE$ C.M. frame versus (a) $M_{\rm miss}(\gamma
\jpsi)$ with $\jpsi \to \ell^+ \ell^-$ reconstructed, (b)
$M_{\rm miss}(\gamma \psp)$ with $\psp \to \pp \jpsi$ reconstructed, and (c)
$M_{\rm miss}(\gamma \psp)$ with $\psp \to \ell^+ \ell^-$ reconstructed.
No evidence for $\jpsi$ or $\psp$ signals can be seen in
the $\gamma \jpsi$ or $\gamma \psp$ missing mass distributions.

\begin{figure*}[htbp]
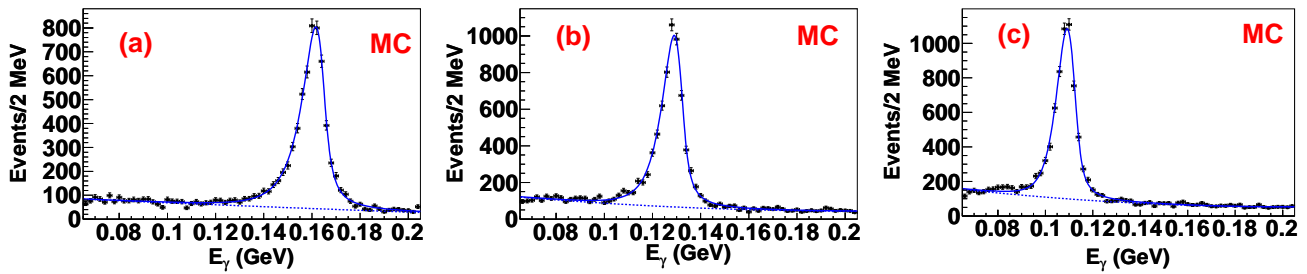

\includegraphics[height=5.5cm,angle=-90]{fig3a.epsi}\hspace{0.2cm}
\includegraphics[height=5.5cm,angle=-90]{fig3b.epsi}\hspace{0.2cm}
\includegraphics[height=5.5cm,angle=-90]{fig3c.epsi}
\caption{The fits to the photon spectra from the $\Upsilon(2S) \to
\gamma \chi_{bJ} \to \gamma \jpsi \jpsi$ MC signal samples with
one $\jpsi$ reconstructed and the $\gamma \jpsi$ recoil mass
within the $\jpsi$ mass region for (a)
$\chi_{b0}$, (b) $\chi_{b1}$ and (c) $\chi_{b2}$, respectively. The $\chi_{bJ}$ shapes are described by
Breit-Wigners convolved with Novosibirsk functions, while Chebychev polynomial functions
are used to describe the background.} \label{one-jpsi-mc}
\end{figure*}

Figure~\ref{one-jpsi-mc} shows the simulated photon spectra in the $\EE$ C.M.
frame from the $\Upsilon(2S) \to \gamma \chi_{bJ} \to \gamma \jpsi
\jpsi$ MC samples. Breit-Wigner (BW) functions convolved with
Novosibirsk functions~\cite{nov} are used as $\chi_{bJ}$ signal shapes while
Chebychev polynomial functions model the combinatorial backgrounds ($\sim11\%$ in
$\chi_{bJ}$ signal region).
The extended maximum likelihood fits to the photon spectra with
all the parameters free are shown in Fig.~\ref{one-jpsi-mc}. Based on the fit results,
the efficiencies are $(5.75\pm 0.12)\%$, $(6.25\pm 0.12)\%$, and
$(5.87\pm 0.12)\%$ for $\Upsilon(2S) \to \gamma \chi_{bJ} \to
\gamma \jpsi \jpsi$ for $J=0$, $1$ and $2$, respectively. Similarly,
the sum of the efficiencies from all the modes is found to be
$(3.40\pm 0.06)\%$, $(3.78\pm 0.06)\%$, and $(3.53\pm 0.06)\%$ for
$\Upsilon(2S) \to \gamma \chi_{bJ} \to \gamma \jpsi \psp$,
$(2.06\pm 0.04)\%$, $(2.15\pm 0.04)\%$, and $(2.09\pm 0.04)\%$ for
$\Upsilon(2S) \to \gamma \chi_{bJ} \to \gamma \psp \psp$ for
$J=0$, $1$ and $2$, respectively.

After all the event selections, no events
from the $\Upsilon(2S)$ MC sample with generic decays survive.
Other possible backgrounds with $\jpsi$/$\psp$ signals from channels
such as $\EE \to \jpsi \chi_{cJ}$, $\psp \chi_{cJ}$, have very small cross-sections
(at the few fb level~\cite{wangkai})
and hence are neglected in the analysis.

\begin{figure}[htbp]
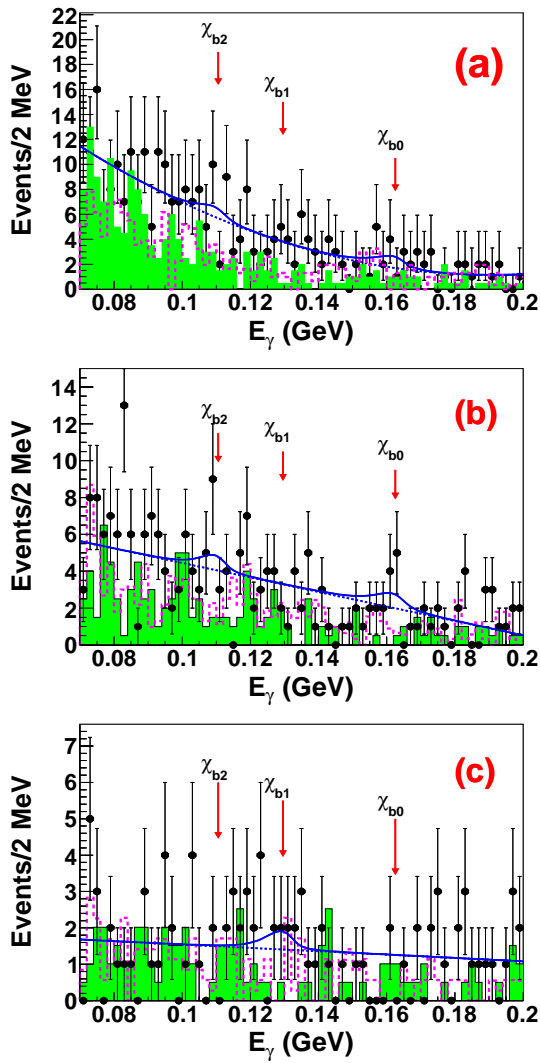

\includegraphics[height=7cm,angle=-90]{fig4a.epsi}\vspace{0.2cm}
\includegraphics[height=7cm,angle=-90]{fig4b.epsi}\vspace{0.2cm}
\includegraphics[height=7cm,angle=-90]{fig4c.epsi}
\caption{The photon spectra in $\Upsilon(2S)$ data for (a) $\gamma
\jpsi \jpsi$, (b) $\gamma \jpsi \psp$, and (c) $\gamma \psp \psp$
final states. The shaded histograms are from normalized
$\jpsi$/$\psp$ mass sidebands events and dashed histograms are
normalized continuum contributions. The fits to the photon spectra
are described in the text. The solid curves are the best fits, the
dashed curves represent the backgrounds. The arrows show the
expected central positions of the $\chi_{bJ}$ states.} \label{one-jpsi-data}
\end{figure}

Figures~\ref{one-jpsi-data}(a), (b) and (c) show the photon
spectra from $\Upsilon(2S)$ data for $\chi_{bJ} \to \jpsi \jpsi$,
$\jpsi \psp$, and $\psp \psp$ candidate events, respectively, with
all the modes included.
Here the shaded histograms show the $\jpsi$ or $\psp$ mass sidebands normalized to the
width of the $\jpsi$ or $\psp$ signal range and the dashed
histograms are the normalized continuum contributions.
The continuum background contribution is extrapolated down to the $\Upsilon(2S)$
resonance. For the extrapolation,
three factors are applied to account for: the relative luminosities of the two
samples, efficiency dependence on the
C.M. energy, and cross section dependence on the C.M. energy. The
cross section extrapolation with C.M. energy is assumed to have a
$1/s$ dependence.

No clear $\chi_{bJ}$ signals are observed in
Fig.~\ref{one-jpsi-data}.
For $\chi_{bJ} \to \jpsi \jpsi$, a unbinned extended maximum
likelihood method is applied to the photon spectrum
with the MC simulated signal shape smeared with a Gaussian function
to take into account a 8.5\% difference in photon energy resolution
between data and MC samples. The photon energy resolution is measured
with $\Upsilon(2S)\to \gamma\chi_{bJ}\to \gamma \gamma \Upsilon(1S)$,
$\Upsilon(1S)\to \mu^+\mu^-$ events.
For $\chi_{bJ} \to \jpsi \psp$ ($\psp
\psp$) decays, an unbinned extended maximum likelihood simultaneous fit is performed to all the modes
 mentioned above. The ratios of the $\chi_{bJ}$ yields in
different modes are fixed to $\eff_i$ ($i$ denotes the $i$-th
mode) with all the intermediate state branching fractions included,
and $\eff_i$ is the MC-determined efficiency for the $i$-th mode.
The fits are performed with the same method as in the $\jpsi \jpsi$ mode.
Figure~\ref{one-jpsi-data} shows the
fit results, where for (b) and (c) the solid curves are the sum of
all the fit contributions, and the dashed curves
are the sum of the background functions.
In all of the modes, the background levels from the fits are a
little higher than the estimations from the normalized continuum
or the normalized $\jpsi$/$\psp$ mass sidebands. It may indicate
that there are double-charmonium production together with one
photon or more particles in $\Upsilon(2S)$ decays.

The upper limit on the number of signal events at the 90\% C.L.
($n^{\rm up}$) is calculated by solving the equation
$\frac{\int_0^{n^{\rm up}}\LK(x)dx} {\int_0^{+\infty}\LK(x)dx} =
0.9$, where $x$ is the number of signal events, and $\LK(x)$ is
the likelihood function depending on $x$ from the fit to the data,
with $x$ being the number of signal events in the fit. The values
of $n^{\rm up}$ are found to be 21, 13, and 22 for $\chi_{bJ} \to
\jpsi \jpsi$; 20, 5.8, and 17 for $\chi_{bJ} \to \jpsi \psp$; and
3.0, 12, and 3.3 for $\chi_{bJ} \to \psp \psp$, for $J=0$, $1$,
and $2$, respectively, when requiring the signal yields to be
non-negative in the fit.


There are several sources of systematic errors for the branching
fraction measurement.
The uncertainty in the tracking efficiency for tracks with angles and
momenta characteristic of signal events is about 0.35\% per track,
and is additive. The photon reconstruction contributes an
additional 3.8\% per photon.
The uncertainty due to particle
identification efficiency is 1.3\% for each pion in $\psp\to \pp
\jpsi$. According to a measurement of the lepton identification
efficiency using a control sample of $\gamma \gamma \to \ell^+
\ell^-$, the MC simulates data within 1.7\% for an
electron-positron pair and 1.7\% for a muon pair.
According to MC simulation, the
trigger efficiency is greater than 99.5\% and we take 0.5\% as
systematic error due to the trigger simulation uncertainty.
Errors on
the branching fractions of the intermediate states are taken from
the PDG~\cite{PDG}, which are
about 12\%, 6.0\% and 5.0\% for $\chi_{b0}$, $\chi_{b1}$ and
$\chi_{b2}$ decays.
By changing the order of the background
polynomial and the range of the fit, the relative difference
in the upper limits of the number of signal events
is 7.6\%-28\% depending
on the decay mode, which is taken as systematic error due
to the uncertainty of fit. For our MC signal samples,
$\jpsi$ and $\psp$ decays are
simulated with a generic decay model. The signal efficiencies are
determined based on the fitted results.
The error on the number of fitted signal events is
less than 2.1\%, which is taken as the MC
statistical error in the efficiency. The masses of $\chi_{bJ}$
have been measured
well~\cite{PDG} and the uncertainties on the masses of $\chi_{bJ}$
do not affect the efficiency determination. Comparing several
theoretical calculations, the maximum values of $\chi_{b0}$ and
$\chi_{b2}$ widths are 2.15~MeV/$c^2$~\cite{gupta} and
0.33~MeV/$c^2$~\cite{laverty}, respectively. The efficiency
differences between these values and the nominal values are
taken as systematic errors due to the uncertainty of resonance
parameters, which are less than 5.2\% and
1.2\% for $\chi_{c0}$ and $\chi_{b2}$ decays.
Finally, the uncertainty on the total number of
$\Upsilon(2S)$ events is 2.3\%. Assuming that all of these
systematic error sources are independent, and combining them in
quadrature, we obtain the total systematic error listed in
Table~\ref{summary}.


Since there is no evidence for signals in the modes studied, we
determine upper limits on the branching fractions of $\chi_{bJ}$
to double charmonia. Table~\ref{summary} lists the upper limits
$n^{\rm up}$ for the numbers of the signal events, detection
efficiencies, systematic errors, and upper
limits on the branching fractions of $\chi_{bJ}$ decays. In order
to calculate conservative upper limits on these branching
fractions, the efficiencies are lowered by a factor of
$1-\sigma_{\rm sys}$ in the calculation.

To summarize, we find no significant signals in the $\chi_{bJ}\to
\jpsi\jpsi$, $\jpsi\psp$, or $\psp\psp$ final states using a
sample of 158 million $\Upsilon(2S)$ events. The results obtained
on the $\chi_{bJ}$ decay branching fractions are listed in
Table~\ref{summary}. Our upper limits are much lower than
the central values predicted in the LC formalism~\cite{vv} and
pQCD calculation~\cite{vv2}, but are consistent
with calculations using the NRQCD factorization
approach~\cite{zhangjuan,sang}.

\begin{table}[htbp]
\caption{Summary of the limits on $\chi_{bJ}$ decays into $\jpsi
\jpsi$, $\jpsi \psp$, and $\psp \psp$. Here $n^{\rm up}$ is the
upper limit on the number of signal events, $\eff$ is the sum of
the efficiencies from different modes with $\jpsi$ and $\psp$
decay branching fractions and trigger efficiency included,
$\sigma_{\rm sys}$ is the total systematic error, and ${\cal B}_R$
is the upper limit on the branching fraction of $\chi_{bJ}$
decays, where the values of
$\BR(\Upsilon(2S) \to \gamma \chi_{bJ})=(3.8\pm0.4)\%, (6.9\pm0.4)\%$ and $(7.15\pm0.35)\%$
for $J=0$, $1$ and $2$ are used~\cite{PDG}.
The upper limits are at 90\% C.L.} \label{summary}
\begin{center}
\begin{tabular}{l  l  c  c  c}
\hline
 Channel & $n^{\rm up}$ &
$\eff$(\%) & $\sigma_{\rm
sys}$(\%)& ${\cal B}_R $ \\
\hline
 $\chi_{b0} \to \jpsi \jpsi$      &  21  & 5.8 & 16 & $7.1\times 10^{-5}$ \\
 $\chi_{b1} \to \jpsi \jpsi$      &  13  & 6.3 & 30  & $2.7\times 10^{-5}$ \\
 $\chi_{b2} \to \jpsi \jpsi$      &  22  & 5.9 & 27  & $4.5\times 10^{-5}$ \\
 $\chi_{b0} \to \jpsi \psp$       &  20  & 3.4  & 17 & $ 1.2\times 10^{-4}$\\
 $\chi_{b1} \to \jpsi \psp$      &   5.8 & 3.8  & 15  & $1.7\times 10^{-5}$\\
 $\chi_{b2} \to \jpsi \psp$      &   17 &  3.5  &  16  & $4.9\times 10^{-5}$\\
 $\chi_{b0} \to \psp \psp$      &    3.0   & 2.1  & 20 & $3.1\times 10^{-5}$\\
 $\chi_{b1} \to \psp \psp$      &    12 &  2.2 &  17  & $6.2\times 10^{-5}$\\
 $\chi_{b2} \to \psp \psp$      &    3.3 &  2.1 & 12  & $1.6\times 10^{-5}$\\
\hline
\end{tabular}
\end{center}
\end{table}


We thank the KEKB group for excellent operation of the
accelerator, the KEK cryogenics group for efficient solenoid
operations, and the KEK computer group and the NII for valuable
computing and SINET4 network support. We acknowledge support from
MEXT, JSPS and Nagoya's TLPRC (Japan); ARC and DIISR (Australia);
NSFC (China); MSMT (Czechia); DST (India); MEST, NRF, NSDC of
KISTI, and WCU (Korea); MNiSW (Poland); MES and RFAAE (Russia);
ARRS (Slovenia); SNSF (Switzerland); NSC and MOE (Taiwan); and DOE
and NSF (USA).


\begin{thebibliography}{**}
\bibitem{chibjdecays} For $\chi_{bJ}$ decays,
MC signal samples are also produced with the helicity amplitude
formulae from $\chi_{cJ} \to VV$ ($V=\omega$ or $\phi$)~\cite{ping}
or from Ref.~\cite{zhangjuan}. The largest difference in the
efficiencies is less than 10\%. The efficiencies with the phase
space mode are lower, which will give conservative upper
limits on the production rates.
\bibitem{nov} The Novosibirsk function is defined as $f(x) =
\exp[-\frac12(\ln^2(1+\Lambda(x-x_0))/\tau^2+\tau^2)]$ with
$\Lambda=$ $\sinh(\tau\sqrt{\ln 4})/(\sigma\sqrt{\ln4})$. The
parameters represent the mean ($x_0$), the width ($\sigma$) and
the tail asymmetry ($\tau$).
\end{thebibliography}
\end{document}